\documentclass[11pt, sf]{iopart}
\usepackage[english]{babel}
\usepackage[T1]{fontenc}
\usepackage{amssymb,amstext}
\usepackage{graphicx}
\usepackage{geometry}%
\usepackage{url}
\usepackage{texdraw}
\usepackage{hyperref}
\usepackage{xspace}
\usepackage[labelfont={sf,bf,footnotesize}, textfont={footnotesize}]{caption}
\usepackage{placeins}

\usepackage{rotating}

\date{\today}

\begin{document}

\title{Evolutionary game dynamics in a growing structured population}

\author{Julia Poncela}

\address{Institute for Biocomputation and Physics of Complex
  Systems (BIFI), University of Zaragoza, E-50009 Zaragoza, Spain}

\author{Jes\'us G\'omez-Garde\~nes}

\address{Institute for Biocomputation and Physics of Complex
  Systems (BIFI), University of Zaragoza, E-50009 Zaragoza, Spain}

\address{Departamento de Matem\'atica Aplicada, ESCET, Universidad Rey
  Juan Carlos, E-28933 M\'ostoles (Madrid), Spain}
  
 \author{Arne Traulsen}

\address{Max Planck Institute for Evolutionary Biology, August-Thienemann-Str.\,2, 
24306 Pl\"on, Germany}

\author{Yamir Moreno}

\address{Institute for Biocomputation and Physics of Complex
Systems (BIFI), University of Zaragoza, Zaragoza 50009, Spain}

\address{Department of Theoretical Physics, University of Zaragoza, Zaragoza 50009, Spain}

\date{\today, \number\hour:\number\minute}

\begin{abstract}
We discuss a model for evolutionary game dynamics in a growing, network-structured population. In our model, new players can either make connections to random preexisting players or preferentially attach to those that have been successful in the past. The latter depends on the dynamics of strategies in the game, which we implement following the so-called Fermi rule such that the limits of weak and strong strategy selection can be explored. Our framework allows to address general evolutionary games. With only two parameters describing the preferential attachment and the intensity of selection, we describe a wide range of network structures and evolutionary scenarios. Our results show that even for moderate payoff preferential attachment, over represented hubs arise. Interestingly, we find that while the networks are growing, high levels of cooperation are attained, but the same network structure does not promote cooperation as a static network. Therefore, the mechanism of payoff preferential attachment is different to those usually invoked to explain the promotion of cooperation in static, already-grown networks.
\end{abstract}

\maketitle

\section{Introduction}

Classical game theory is a branch of applied mathematics that has been developed to describe strategic interaction between fully rational individuals \cite{neumann:1944ef}. Evolutionary game theory is an elegant way to abandon the often problematic rationality assumption of classical game theory and to introduce a natural dynamics to that classical concept \cite{maynard-smith:1973to,hofbauer:1998mm}. In the past, evolutionary game theory has been used to describe either cultural learning dynamics 
or genetic reproduction under frequency dependent selection \cite{nowak:2004aa}.
More recently, it has attracted a lot of interest in the physics community in the context of nonlinear dynamics \cite{sato:2002le,sato:2003le}, disordered systems \cite{berg:1998aa,vainstein:2001hp,galla:2007aa}, finite size effects \cite{helbing:1993aa,traulsen:2006hp}, or spatially extended systems \cite{lindgren:1994to,szabo:1998wv,abramson:2001nx,schweitzer:2002le,santos:2005pm,perc:2006aa,perc:2008qd}. Statistical mechanics provides a powerful tool to describe evolutionary game dynamics in spatially extended, structured populations. Besides, in the last decade network theory has contributed significantly to our quantitative understanding of structured systems which go beyond the regularity of simple lattices \cite{boccaletti:2000aa}.

A typical setup is the following: Agents are assigned to the nodes of a network, which can be a regular lattice or have a more complex structure. Then, agents play an evolutionary game in which more successful strategies spread on the system. Describing these systems analytically is tedious and only possible in special cases \cite{hauert:2002mn,ohtsuki:2006na,szabo:2007aa,floria:2009aa}. Moreover, there are few general statements that can be made on evolutionary dynamics in such spatial systems \cite{sanchez:2009aa}. 

Here, we drop another simplifying assumption and consider evolutionary games in growing, network-structured populations. In other words, instead of taking a growth algorithm for a particular network and later simulate evolutionary dynamics on that network, we grow the network while the evolutionary game is played. The interplay between growth and evolutionary game dynamics leads to interesting network structures and allows to disentangle effects based on topology from effects based on growth of the network.

\section{Growing structured populations}

We address the case of a growing population in which new individuals establish connections to the existing individuals, see also \cite{poncela:2008aa}. 
The newcomers can either connect to $m$ arbitrary individuals or preferentially attach to those that have been successful players in the past. Success is based on the cumulated payoff $\pi$ from an evolutionary game, which each individual plays with all its neighbors on the network. For the model itself, we do not have to specify the kind of the game or the number of strategies. 

We start from a small complete network of $N_0$ individuals of one strategy. Subsequently, new individuals arrive and form connections to existing individuals. 
Evolutionary dynamics proceeds in the following way: 
\begin{itemize}
\item[(i)] 
In each time step,
every individual $j$ 
plays with all its neighbors and obtains an accumulated payoff $\pi_{j}$.

All players chose between their old strategy and the strategy of a randomly selected neighbor synchronously.  Player $j$ will adopt the strategy of its randomly selected neighbor $i$ with probability 
\begin{equation}
T_{j \to i} = \frac{e^{+ \beta \cdot \pi_i}}{\e^{+ \beta \cdot  \pi_i} + \e^{+ \beta \cdot  \pi_j}},
\end{equation}
where $\beta$ is the intensity of selection. With probability $1 - T_{j \to i} $, it will
stick to its old strategy.  
For $\beta \ll 1$, selection is weak and the game is only a linear correction to random strategy choice. 
For strong selection, $\beta \to \infty$, it will always adopt a better strategy and it will never adopt a worse strategy. 
This process is routinely used in evolutionary game dynamics \cite{szabo:1998wv,blume:1993jf,traulsen:2007cc}.

\item[(ii)] 
Every $\tau$ time steps, a new individual with a random strategy is added to the system.  
For $\tau \ll 1$, several nodes are added before individuals change strategies. 
For $\tau \gg 1$, the network grows very slowly and the game dynamics can bring the system close to 
equilibrium before a new node is added. 
The new individual establishes $m$ links to preexisting nodes, which are chosen preferentially according to their performance in the game in the last time step.
Node $j$ is chosen as an interaction partner with probability
\begin{equation}
p_j = \frac{e^{+ \alpha \cdot \pi_j}}{\sum_{l=1}^N e^{+ \alpha \cdot  \pi_l}},
\label{ppatt}
\end{equation}
where $N$ is the number of nodes that already exist when the new node is added. 
The remaining $m-1$ links are added in the same way, excluding double links. 
For $\alpha=0$, the newcomer attaches to a randomly chosen existing node.  
For small $\alpha$, attachment is approximately linear with payoff.
For high $\alpha$, the newcomers will make connections to only very few 
nodes with high payoffs.
For $\alpha \to \infty$, all newcomers will always attach to the $m$ most successful
players. 

\end{itemize}

Since $m$ links and a single node are added in each $\tau$ time steps, the average degree of the network is given by 
\begin{equation}
\frac{  N_0(N_0-1) \frac{1}{2} + m \frac{t}{\tau} }{N_0 +  \frac{t}{\tau}} ,
\end{equation}
where $t$ is the number of time steps that has passed. Throughout this work, we will concentrate on $m=2$ and $N_{0}=3$. 

Let us first focus on the simplest case in which each interaction leads to the same payoff, which we set to one.
Then, the payoffs $\pi_j$ are just the number of interactions an individual has, i.e.\ the degree $\kappa_j$ of the node (normalizing by the degree of the node would essentially wash out the effect of the topology at this point \cite{santos:2006je,pusch:2008pb}).
Evolutionary dynamics of strategies has no consequences and thus, the topology is independent of $\beta$. 
This allows us to discuss the growth dynamics without any complications arising from the dynamics of strategies. 
We have several simple limiting cases:

\begin{itemize}
\item 
For $\alpha = 0$, the newcomer attaches at random to a new node. This leads to a network in
which the probability that a node has $k$ links decays exponentially fast with k. The situation correspond to the case studied in \cite{barabasi:1999aa}, as individuals introduced earlier are likely to get more links. In this case, topology is independent of strategies for all intensities of selection $\beta$
even when individuals play different strategies leading to different payoffs. 
Whenever $\alpha>0$, there is an interplay between topological dynamics and strategy dynamics. 

\item 
For $\alpha \ll 1$, we can linearize $p_j$. 
In this case, we obtain 
\begin{equation}
p_j = \frac{\alpha^{-1}+ \kappa_j}{\sum_{k=1}^N {(\alpha^{-1} +  \kappa_k)}}.
\end{equation}
Thus, we recover the linear preferential attachment model introduced by Dorogovtsev et al. \cite{dorogovtsev:2000wl}. When strategies differ in their payoffs, then not only the degree, but also the strategy of the nodes and their neighbors will influence the probability to attach to a node. 

\item
When $\alpha$ is large, we will typically observe a network 
in which $m$ of the $N_0$ nodes of the initial complete network will be connected to almost all nodes that have been added during the growth stage. The emergence of these super-hubs hinges on the nonlinearity in Eq. \ref{ppatt}. 

\end{itemize}

\begin{figure}
\begin{center}
\includegraphics[width=0.9\textwidth,angle=0]{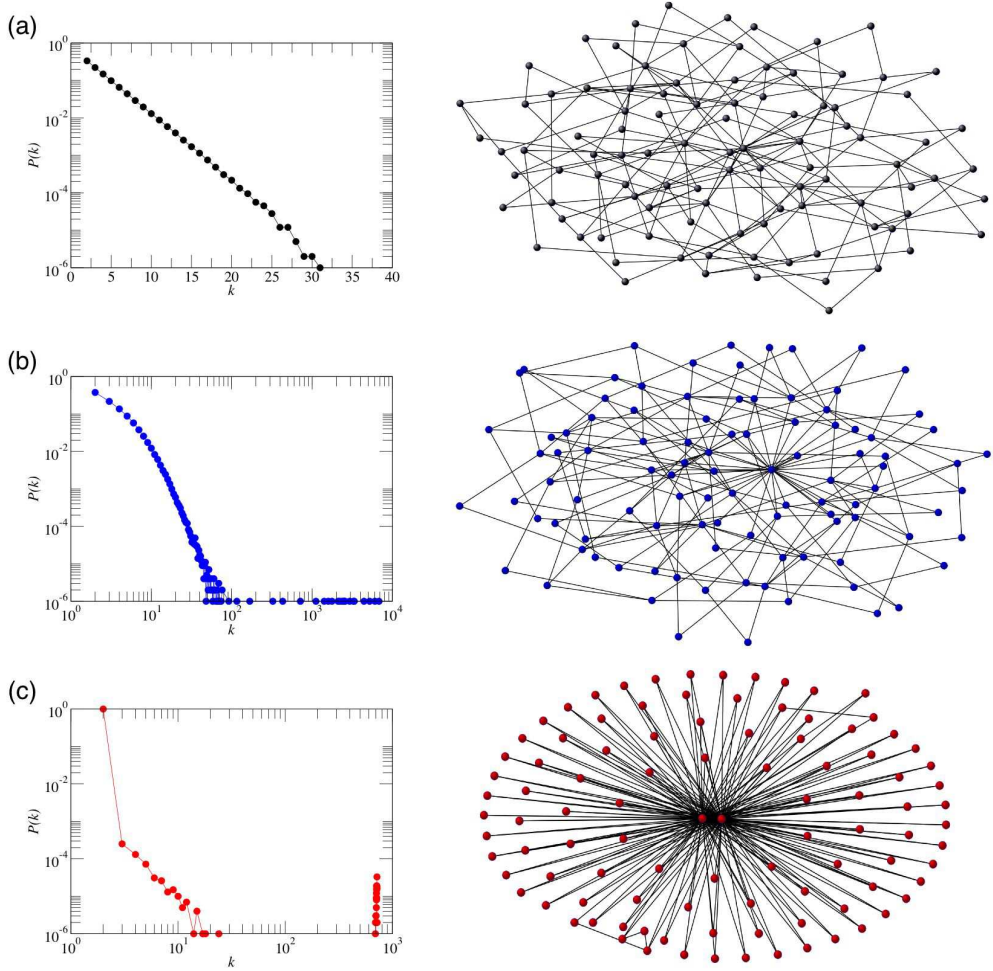}
\caption{Networks for a game in which both strategies have identical payoffs, such that the payoff is given by the degree of a node. The left hand side shows the degree distributions of networks of size $N=10^4$, while the right hand side are snapshots of networks of $N=100$ nodes. 
(a) For $\alpha = 0.0$, the degree distribution decays exponentially.  
(b) For $\alpha = 0.1$, some highly connected nodes appear in the network and
the degree distribution begins to resemble a power-law. 
(c) Already for $\alpha = 1.0$, the vast majority of nodes (>99.9 \%) has only two links. In addition, $m=2$ of the $N_0=3$ initial
nodes are connected to almost all other nodes
(degree distributions are obtained from an average over 100 networks, 
note that the $x$-axis is linear in (a), but logarithmic in (b) and (c)).
}
\label{figA}
\end{center}
\end{figure}

Examples for the network structures in these limiting cases are given in Fig.~\ref{figA}. 
Next, we turn to evolutionary games in which the payoff per interaction is no longer constant,
but depends on the strategies of the two interacting individuals. 
In general, such an interplay of evolutionary dynamics of the strategies and the payoff-preferential attachment will change the structure of the network.

\section{Playing evolutionary games during growth}

In principle, our framework allows to address any game between individuals, even repeated games or games with many strategies can be considered. However, we focus on the Prisoner's Dilemma here as an example of a one-shot game with two strategies \cite{rapoport:1965pd,axelrod:1984yo,nowak:2006pw}. Two players can choose between cooperation and defection. In the simplest case, there is a cost $c$ to cooperation, whereas a cooperative act from an interaction partner leads to a benefit $b\, (>c)$. The game can be written in the form of a payoff matrix, 
\begin{equation}\label{eq:Pmatrix}
\bordermatrix{
  & C & D \cr
C & b-c & -c \cr
D & b & 0 \cr}.
\end{equation}
No matter what the opponent does, defection leads to a higher payoff (due to $b>b-c$ and $0>-c$).
Thus selfish, rational players should defect. 
Similarly, if the payoff determines reproductive fitness, evolution will lead to the spread of defection. 
However, the payoff for mutual defection is smaller than the payoff for mutual cooperation ($b-c>0$)
and thus players face a dilemma.
One way to resolve the dilemma is to consider structured populations in which players only interact with 
their neighbors \cite{nowak:1992pw}. Here, we follow this line of research and consider in addition growing populations, as discussed above.

\begin{figure}
\begin{center}
\includegraphics[width=0.6\textwidth,angle=-90]{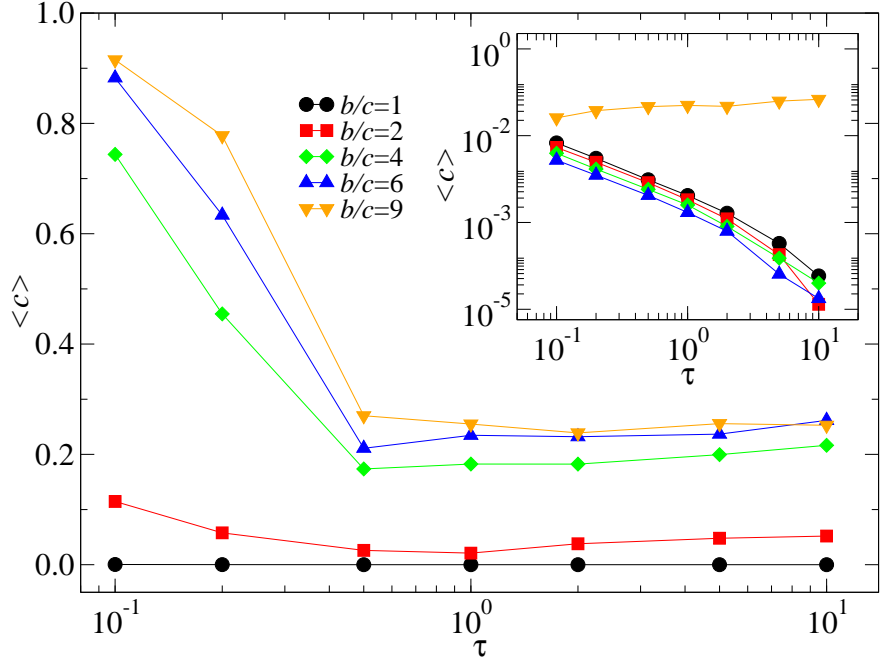}
\caption{The average level of cooperation under strong selection ({$\beta=1$}) and $\alpha=1$
depending on the time scale of attachment, $\tau$. 
Cooperation benefits most from small values of $\tau$, i.e.\ when many new nodes are added before 
players update their strategies.
For random attachment ($\alpha=0$, inset) cooperation does not emerge, only for high benefit to cost ratios a few cooperators prevail 
($m=2$, $N_0=3$, values obtained from $10^2$ averages over networks of final size $N=1000$, averaged when the network stops growing). }
\label{figB}
\end{center}
\end{figure}

Typically, one is interested in the promotion of cooperation on different network structures. 
Fig.\ \ref{figB} shows the average level of cooperation for strong selection as a function of $\tau$. It turns out that payoff preferential attachment increases the level of cooperation significantly compared to random attachment. This effect is also present for weak selection, but less pronounced.
Cooperation increases most for small $\tau$, i.e.\ when many nodes are added before strategies are changed. 
This puts the system further from equilibrium, whereas the case of large $\tau$ means that strategies have been equilibrated at least locally
before the next new individual with a random strategy is added to the system. Note that for $\tau$ larger than a certain value, cooperation levels become independent of $\tau$, which points out that playing once a given number of new players are incorporated is enough to reach a dynamical equilibrium.

\begin{figure}
\begin{center}
\includegraphics[width=0.9\textwidth,angle=0]{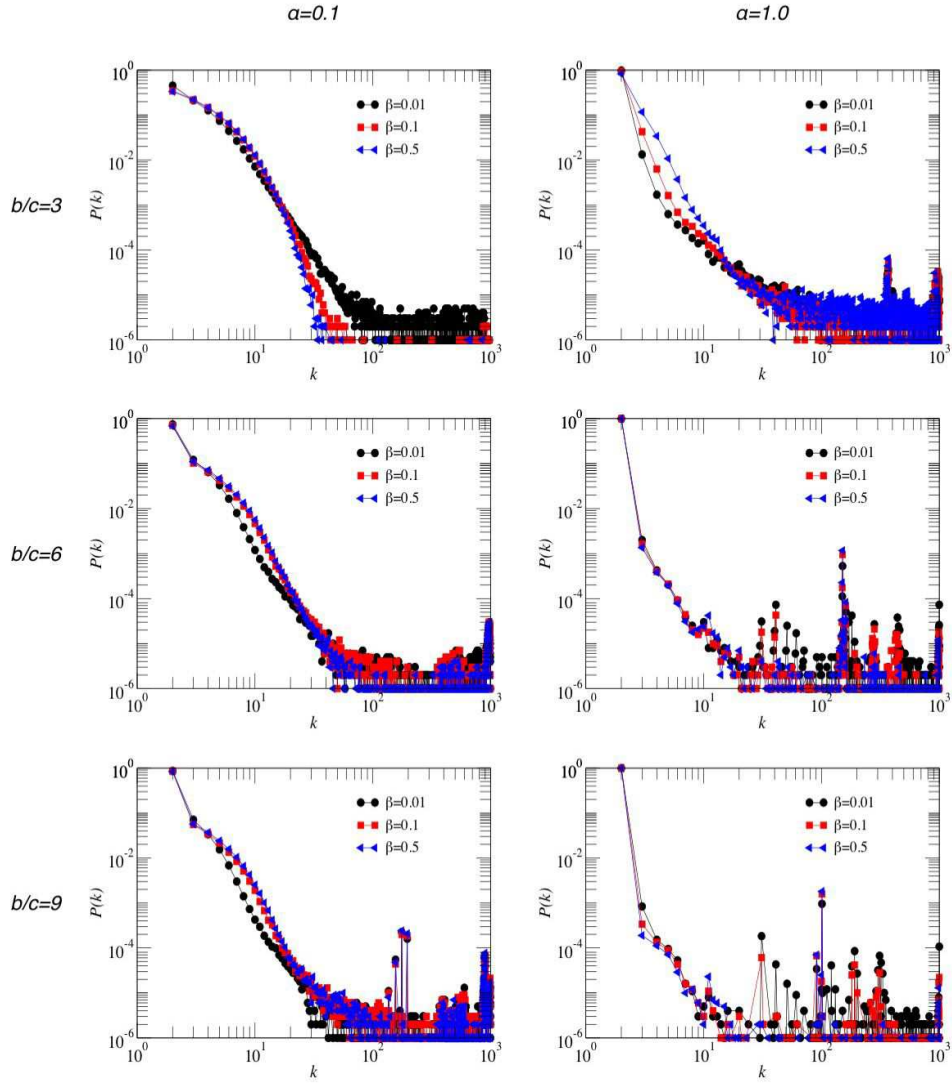}
\caption{Impact of the game dynamics on the degree distribution at the end of network growth. Left column corresponds to $\alpha=0.1$, while the right one is for $\alpha=1$. In general, game dynamics has only a weak impact on the topology of the system. However, there is a trend that stronger selection increases the number of nodes with
fewer links and decreases the number of highly connected nodes
($N_{0}=3$, $m=2$, $\tau=0.1$, 
distributions averaged $10^3$ over realizations of networks of $10^3$ nodes each).}
\label{figC}
\end{center}
\end{figure}

Since there is an interaction between strategy dynamics and network growth, the topology will change under selection. In Fig. \ref{figC}, we show how the topology for the Prisoner's Dilemma changes with the benefit to cost ratio $b/c$, the intensity of selection $\beta$ and the attachment parameter $\alpha$ (see also Fig.\ref{figA}). It turns out that the influence of the game on the degree distribution is relatively weak, for small degrees a clear  difference is only found for large $\alpha$ and small $b/c$. The distribution of the relatively few nodes with many connections, however, is more sensitive to changing either $b/c$ or $\beta$. 

\section{Promotion of cooperation in growing networks}

\begin{figure}
\begin{center}
\includegraphics[width=0.35\textwidth,angle=-90]{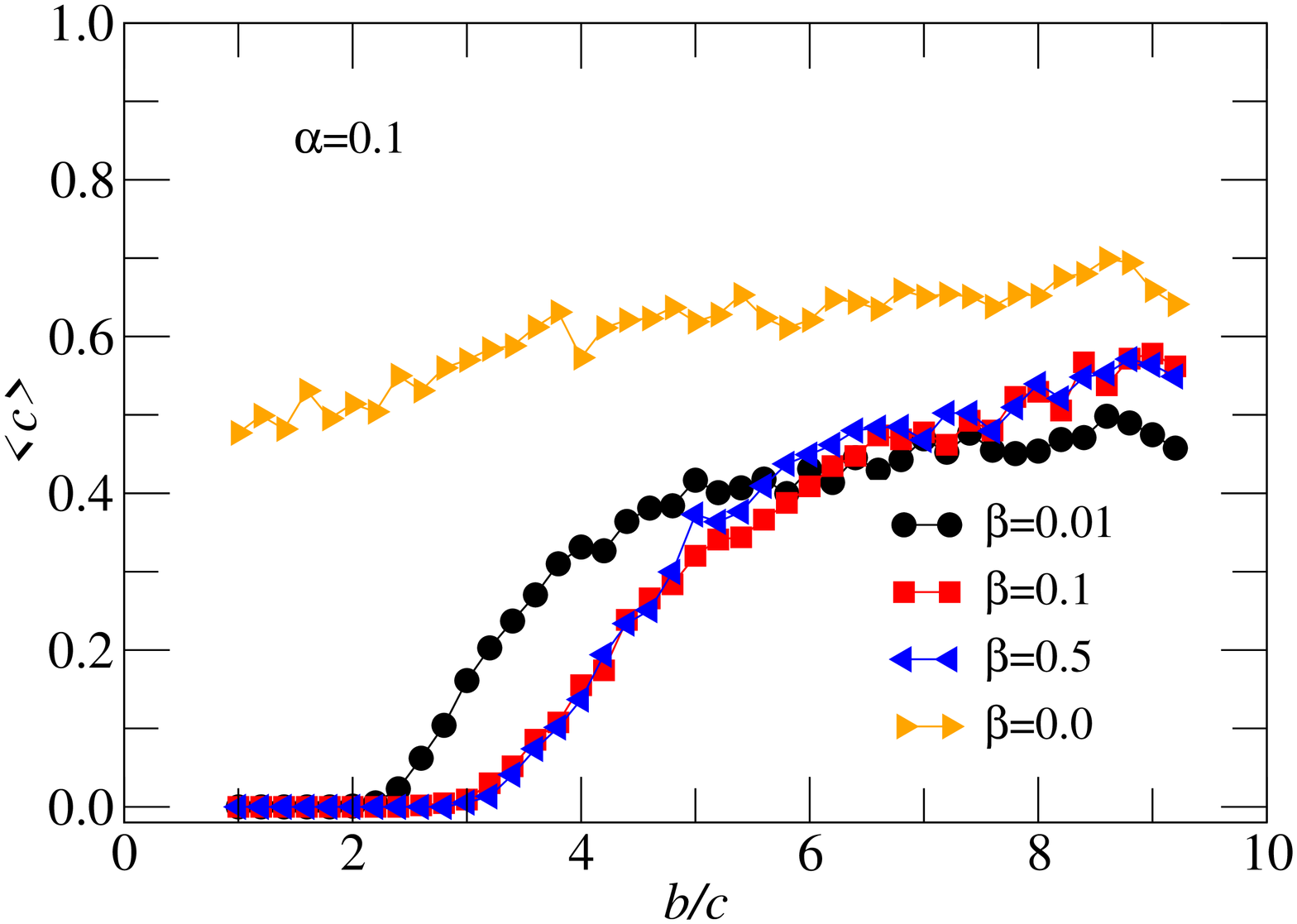}
\includegraphics[width=0.35\textwidth,angle=-90]{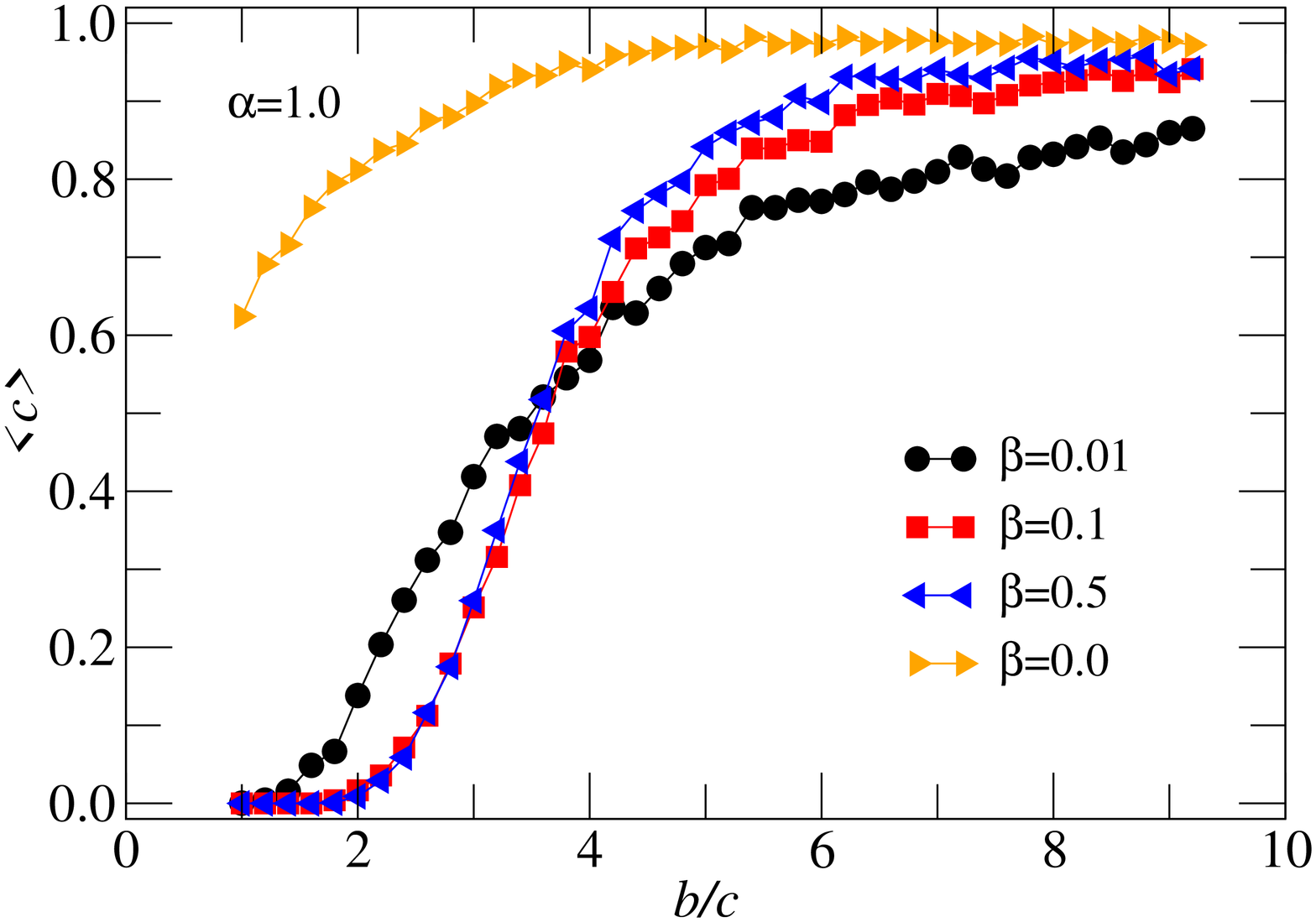}
\caption{The average level of cooperation $\langle c \rangle$ $10^4$ time steps after the network stops growing.
For $\alpha = 0.1$ (left) the level of cooperation exceeds 50 \% only for very high benefit to cost ratios $b/c$. 
For $\alpha = 1.0$ (right), the abundance of cooperators is significantly higher.
Even for neutral strategy dynamics ($\beta=0$), payoff preferential attachment can lead to high levels of cooperation in this case
($N_{0}=3$, $m=2$, $\tau=0.1$,
averages over $10^3$ different networks of size $10^3$).}
\label{figD1}
\end{center}
\end{figure}

As in most structured populations, cooperators that are disadvantageous in 
the Prisoner's Dilemma in
well-mixed population
benefit from the spatial structure. Of course, this effect is larger when cooperation becomes more
profitable, i.e.\ when the benefit to cost ratio $b/c$ increases. 
It turns out that for weak payoff preferential attachment (small $\alpha$), the promotion of cooperation is relatively weak
and levels of cooperation beyond 50 \% are only reached when cooperation is very profitable, see Fig. \ref{figD1}.
However, when the probability to attach to the most successful nodes becomes large (large $\alpha$), then
the average fraction of cooperators becomes larger, approaching one when the benefit cost ratio $b/c$ is large. 

Interestingly, for small $b/c$ ratios, the abundance of cooperators decreases with increasing $\beta$,
whereas it increases with the intensity of selection for large $b/c$ ratios. 
The existence of a threshold for intermediate $b/c$ can be illustrated as follows for large $\alpha$: 
Assume that we start from $N_{0}$ fully connected cooperator nodes. For 
$\tau<1$, we add $1/\tau$ nodes with $m=2$ links, on average
half of them defectors and half of them cooperators. 
All new players interact only with the initial cooperator nodes, 
such that an initial cooperator will on average obtain
$\frac{m}{N_{0} \tau}$ new links. 
Thus, the payoff of a new defector is $m b$. 
The average payoff of an initial cooperator is 
$(b-c)(N_{0}-1+\frac{1}{2} \frac{m}{N_{0} \tau})  - c \frac{1}{2} \frac{m}{N_{0} \tau}$. 
Both payoffs are identical for 
\begin{equation}
\frac{b}{c} = \frac{
\frac{1}{\tau}+ \frac{N_{0}(N_{0}-1)}{m}
}{
\frac{1}{2\tau}-N_{0}+ \frac{N_{0}(N_{0}-1)}{m}
}.
\end{equation}
For large $b/c$, cooperators will dominate in the very beginning of network growth. 
The threshold increases with $\tau$ and decreases with $N_{0}$: The larger the initial cooperator
cluster and the more nodes are added before strategies are updated, the easier it is for cooperation
to spread initially. This argument shows qualitatively that a crossover 
in the abundance of cooperators
should exist, and therefore that above a certain threshold, it is easier for cooperation to spread. Only in the very beginning of network growth, this argument will hold quantitatively.

\begin{figure}
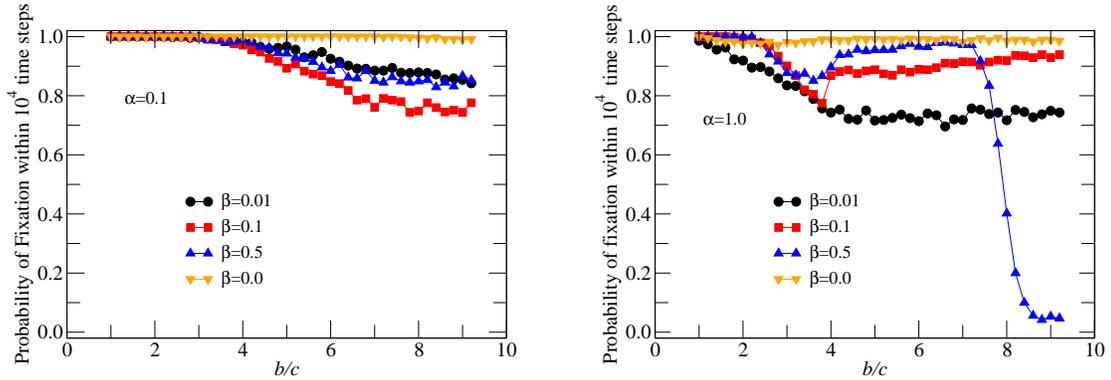

\begin{center}
\vspace{0.5cm}
\includegraphics[width=0.45\textwidth,angle=0]{Fig5a.eps}
\qquad
\includegraphics[width=0.45\textwidth,angle=0]{Fig5b.eps}
\caption{The probability of fixation for one strategy within $10^4$ time steps after growth has stopped in dependence of the attachment parameter $\alpha$ (left column $\alpha=0.1$, right column $\alpha=1$) for different intensities of selection
$\beta$. 
For small $\alpha$, the degree distribution decays exponentially and fixation is relatively fast, regardless of the intensity of selection. For $\alpha=1.0$, the network is more heterogeneous. As discussed in the text, for intermediate values of $b/c$ ($\approx 3.5$), the probability of fixation within $10^4$ time steps is smaller than for higher and smaller $b/c$. 
For very high $b/c$ and strong selection, one observes a coexistence of cooperators and defectors for a very long time rather than fixation for one of the strategies
($N_{0}=3$, $m=2$, $\tau=0.1$, averages over $10^3$ independent realizations of a network of $10^3$ nodes).}
\label{figD2}
\end{center}
\end{figure}

In general, the average level of cooperation can be based on two very different scenarios: 
Either it is the fraction of realizations of the process that ultimately ends in full cooperation, 
or it is the average abundance of cooperators in a network in which both cooperators
and defectors are present. 
For any finite intensity of selection $\beta$, we have $T_{j \to i}>0$, regardless of the payoffs. Thus, after growth has stopped, our dynamics describes a recurrent Markov chain with two absorbing states in which all players follow one of the two strategies. Therefore, ultimately one of the two strategies will go extinct, in contrast to evolutionary processes that do not allow disadvantageous strategies to spread \cite{poncela:2008aa}. 
However, the time to extinction can become very large, in particular when the intensity of selection is high or the population size is large \cite{traulsen:2007cc,antal:2006aa}.
In Fig. \ref{figD2}, we analyze this issue 
numerically. We address the probability that fixation (for either cooperation or defection) occurs within $10^4$ time steps after the network has stopped growing. 
For small $\alpha$, the results follow the intuition from well-mixed populations: Fixation within this time is more likely
if the intensity of selection is weaker. With increasing benefit to cost ratio, fixation times increase and a fixation within the first $10^4$ time steps becomes less and less likely.

For large $\alpha$, however, fixation is faster for strong selection (large $\beta$) for a wide range of parameters. Only when the $b/c$ ratio is very high, fixation times are very large under strong selection.  
This is based on the peculiar structure of the network obtained for large $\alpha$. In addition, 
we observe an area in Fig.\ \ref{figD2} where the fixation time increases slightly before it 
decreases again, i.e.\ the probability for fixation in the first $10^4$ time steps has a minimum. 
Interestingly, this occurs for the range of $b/c$ ratios where the average levels of cooperation intersect at 50 \% for the different intensities of selection. In this parameter region, neither cooperators nor defectors are clearly favored. 
Thus, they will initially both spread. 
When the abundance of both strategies is approximately constant in the beginning, then it will be more difficult to completely wipe out one strategy later. Thus, the increased time of fixation in the parameter region where 
the abundance of cooperation becomes $50 \%$ makes intuitive sense. 

\begin{figure}
\begin{center}
\vspace{0.5cm}
\includegraphics[width=\textwidth,angle=0]{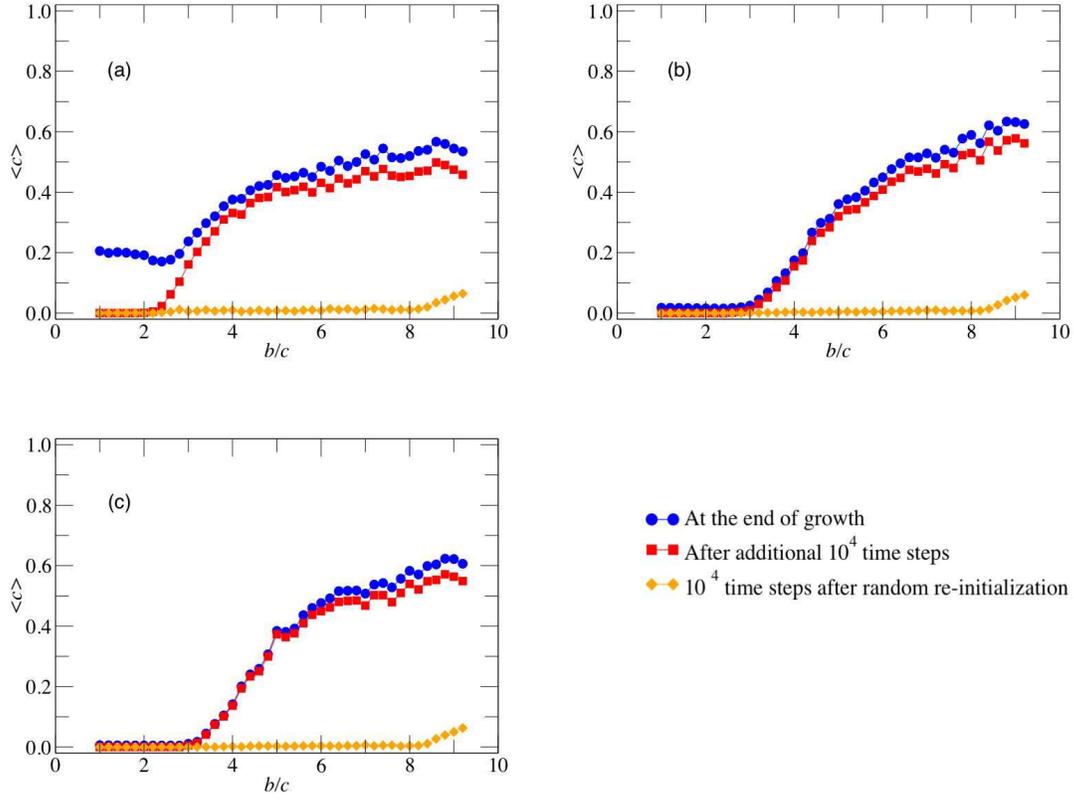}
\caption{Does cooperation benefit from the topology only or also from growth?
Here we analyze the average level of cooperation in three cases, 
(i) once the network is fully grown
(ii) after the game dynamics has proceeded $10^4$ additional steps beyond the growth phase of the network
(iii)  $10^4$ time steps after the fully grown network has been re-initialized with random strategies.
Clearly, the grown networks itself do not promote cooperation significantly. Instead, the growth phase is of crucial importance. The intensity of selection has only a minor influence on the phenomenon. (a) $\beta = 0.01$, (b) $\beta=0.1$, and (c) $\beta=0.5$
($N_{0}=3$, $m=2$, $\tau=0.1$, 
averages over at least $10^2$ networks of size $10^3$, $\alpha=0.1$ in all cases).
}
\label{figE}
\end{center}
\end{figure}

\section{Does cooperation benefit from growth or only from topology?}

Typically, the promotion of cooperation in the Prisoner's Dilemma is analyzed on static networks. 
Our model allows a feedback between the game dynamics and the growth of the network. 

What happens when the network stops growing? Typically, one would expect that defectors 
profit from growth, because there is a steady flow of new cooperators that they can potentially exploit.
Thus, cooperation should increase if the game dynamics proceeds on the fully grown, static network. 
This has also been observed in a previous paper \cite{poncela:2008aa}. In contrast to that paper,
here we have changed the game dynamics in such a way that individuals sometimes can also
adopt a worse strategy. 
It has been shown that this seemingly small change can significantly decrease the level of cooperation
\cite{ohtsuki:2006rg}.
The overall level of cooperation drops significantly and is only higher than 50\% if cooperation is
very profitable. 
In addition, the level of cooperation now decays once the network no longer grows, see Fig. \ref{figE}. 
This means that cooperators, not defectors, benefit from the continuous supply of new players.

Next, we can ask whether the topologies that are obtained from the network growth are
powerful promoters of cooperation at all. This can be tested by taking the fully grown, static network
and run the game dynamics on the fixed network with initially random strategies, 50\% cooperators
and 50 \% defectors. Interestingly, this does not lead to any significant levels of cooperation, cf.\ Fig. \ref{figE}. 
Thus, our model of network growth based on payoff preferential attachment itself leads to comparably high levels of cooperation, while the resulting topology alone does
not support cooperation in the Prisoner's dilemma. 

\section{Discussion}

Our model for evolutionary game dynamics in a growing, network-structured population
is a dynamical network model \cite{gross:2008aa}. 
Here, the network grows,
in contrast to most models for evolutionary games
on dynamical networks that consider a constant population size \cite{skyrms:2000sk,ebel:2002aa,ebel:2002cc,zimmermann:2005aa,pacheco:2006pa,pacheco:2006pb,santos:2006aa,biely:2007aa,pacheco:2008aa,van-segbroeck:2008px,van-segbroeck:2009mi,gomez-gardenes:2007oy,gomez-gardenes:2008ez}. 
Individuals cannot break links and cannot control directly how many new individuals will establish connections with them. 

An important difference with previous work \cite{poncela:2008aa} is that under strong payoff preferential attachment, the topology of the networks generated are dominated by the presence of a few hubs, which attract most of the links of the rest of the nodes. The existence of very few hubs and a large number of lowly connected nodes in network models have been previously noticed \cite{leyvraz:2002fi}. In fact, it has been shown that when networks are grown following a non-linear preferential attachment rule of the sort $p_j=\frac{k_j^{\nu}}{\sum_{l=1}^N k^{\nu}_l}$, with $\nu > 1$, star like structures are obtained \cite{krapivsky:2000kl}. Here, we have shown that the same kind of networks are produced when the dynamics driving the attachment process is dominated by the most successful players. 
Even when payoff preferential attachment is not too strong
(for instance, for $\alpha=0.1$), super-hubs emerge, a clear mark that successful players are likely to attract many of the links of the new nodes.

If newcomers preferentially attach to the successful players in the game, then high levels of cooperation are possible. 
But this cooperation hinges upon the growth of the network, the population structure alone would not lead to such high levels of cooperation. Thus, payoff preferential attachment differs from the usual promotion of cooperation in structured populations. In particular, it has been suspected 
that heterogeneous structures favor cooperative behavior due to the existence of hubs. However, as Fig. \ref{figE} shows, the presence of super-hubs is not enough to sustain cooperation in the networks grown following the scheme discussed here.

In other models, the probability to adopt a strategy that performs worse is zero \cite{santos:2005pm,poncela:2008aa,floria:2009aa}. In particular together with synchronous updating of strategies, this can lead to evolutionary deadlocks, i.e. situations in which both strategies stably coexist. 
Here, we have adopted an update scheme in which individuals sometimes adopt a strategy 
that performs worse. Due to the presence of such irregular moves, sooner or later (often much later) one
strategy will reach fixation. However, when $\beta$ and the ratio $b/c$ are large enough, both cooperation and defection can coexist for a long time. 

Let us also remark that our growth mechanism has also another interesting feature: 
It has been shown that the average level of cooperation obtained in static, scale-free networks, is robust to a wide range of initial conditions \cite{poncela:2007aa}. However, for the networks grown using the payoff preferential attachment, the initial average number of cooperators in the neighborhood of the super-hubs determines the fate of cooperation in the whole network, leading to a much more sensible dependence on the initial state of the system.
From this point of view, the weak dependence on the initial conditions 
reported in static scale-free networks is not trivial. 

Finally, we point out that it would be of further interest to study the model discussed here with other $2 \times 2$ games. As we have shown, the game dynamics seems to have a weak impact on the structure of the resulting networks. Whether or not this holds in general will elucidate the question of the influence of different games on the network formation process. For instance, within the model discussed in \cite{santos:2006aa}, different topologies emerge when different game dynamics are implemented.

In summary, our model shows that the interplay of game dynamics and network growth leads to complex network structures. 
Moreover, not only the structure of the interaction network is important for the evolution of cooperation, but also the particular way this structure is obtained. Our work shows that playing while growing can lead to radically different results with respect to the most studied cases in which game dynamics proceeds in static networks.

\section*{Acknowledgements}
We thank J.M. Pacheco for valuable comments. Y. M. thanks the hospitality of the Max Planck Institute for Evolutionary Biology, where parts of this work have been finished. 
We gratefully acknowledge funding by the COST action ``Physics of Conflict and Cooperation'' (A.T. and Y.M.), 
the Emmy-Noether Program of the DFG (A.T.), the MICINN (Spain) through
Grants FIS2006-12781-C02-01, and FIS2008-01240. Y.M. is supported by MICINN (Spain) through the Ram\'on y Cajal Program. 

\section*{References}


\end{document}